\documentclass{article}

\usepackage{arxiv}

\usepackage[utf8]{inputenc} 
\usepackage[T1]{fontenc}    

\usepackage{url}            
\usepackage{booktabs}       
\usepackage{amsfonts}       
\usepackage{nicefrac}       
\usepackage{microtype}      
\usepackage{lipsum}
\usepackage{graphicx}
\graphicspath{ {./images/} }
\RequirePackage{amsthm,amsmath,amsfonts,amssymb}
\RequirePackage[numbers,sort&compress]{natbib}
\RequirePackage[colorlinks,citecolor=blue,urlcolor=blue]{hyperref}
\RequirePackage{graphicx}
\usepackage{algorithmic}
\usepackage{algorithm}

\theoremstyle{plain}

\newtheorem{theorem}{Theorem}[section]
\newtheorem{lemma}[theorem]{Lemma}
\theoremstyle{definition}

\newtheorem*{remark}{Remark}

\title{When Is Causal Inference Possible? A Statistical Test for Unmeasured Confounding}

\author{
 Muye Liu \\
  Department of Statistics\\
  Purdue University\\
  West Lafayette, IN 47907 \\
  \texttt{liu2546@purdue.edu} \\
   \And
 Jun Xie \\
  Department of Statistics\\
  Purdue University\\
  West Lafayette, IN 47907 \\
  \texttt{junxie@purdue.edu} \\
}

\begin{document}
\maketitle
\begin{abstract}
This paper clarifies a fundamental difference between causal inference and traditional statistical inference by formalizing a mathematical distinction between their respective parameters. We connect two major approaches to causal inference, the potential outcomes framework and causal structure graphs, which are typically studied separately. While the unconfoundedness assumption in the potential outcomes framework cannot be assessed from an observational dataset alone, causal structure graphs help explain when causal effects are identifiable through graphical models. We propose a statistical test to assess the unconfoundedness assumption, equivalent to the absence of unmeasured confounding, by comparing two datasets: a randomized controlled trial and an observational study. The test controls the Type I error probability, and we analyze its power under linear models. Our approach provides a practical method to evaluate when real-world data are suitable for causal inference.
\end{abstract}


\section{Introduction}

Causal inference seeks to uncover cause-effect relationships from experimental or observational data. Unlike standard statistical inference, which focuses on probability distributions, causal inference relies on the notion of intervention \cite{pearl2009causality,spirtes2000causation}. Statistical experiments, such as randomized controlled trials (RCTs), involve actual interventions and serve as the primary data source for causal inference. In observational settings, causal inference can still be possible when certain conditions allow adjustment to remove non-causal dependencies, or when interventions can be approximated using statistical methods \cite{li2025featurematchinginterventionleveraging}. This paper aims to assess the conditions under which causal inference is possible from observational data.

There are two fundamental approaches to causal inference: the potential outcomes
framework \cite{splawa1990application,rubin1977}, developed within the statistical community, and causal structure graphs \cite{pearl2009causality}, primarily advanced in the
computer science community. In statistics, and under the potential outcomes framework, the primary focus has traditionally been on estimating treatment effects, that is, assessing whether a treatment causes changes in an outcome of interest. Causal effect estimation has historically relied on the design and analysis of RCTs,  dating back to the seminal contributions of Fisher and Neyman \cite{splawa1990application,fisher1937design}. Through randomization, an RCT implements a perfect intervention \cite{scholkopf2021toward} on the treatment variable (treatment assignment), allowing causal effects to be identified. However, RCTs are often expensive, logistically challenging, or ethically infeasible. As a result, observational studies are more common in many fields, but they require additional assumptions to make causal inference.

The most prominent of these is the unconfoundedness assumption \cite{rosenbaum1983central}, also known as the ignorability or conditional exchangeability assumption. It is mathematically formulated within the potential outcomes framework. The potential outcomes framework defines that each experimental unit has two potential outcomes: one if exposed to the control treatment and one if exposed to the active treatment. Under the unconfoundedness assumption, causal effects can be estimated using methods such as regression or machine learning (details in Section 2.2). However, because we only observe one potential outcome per individual but never both, such type of assumptions based on potential outcomes are inherently untestable from observational data alone.

Dropping the unconfoundedness assumption means causal effects are no longer identifiable, and no general solution exists for this case. Some identification strategies have been proposed, including instrumental variables \cite{angrist1990lifetime, angrist1996identification,imbens2014instrumental,andrews2019weak} and sensitivity analysis \cite{manski1990nonparametric, rosenbaum1983assessing, rosenbaum2002observational}; the latter provides an upper bound on confounding effects. Our primary goal here is to develop a statistical test to assess whether the unconfoundedness assumption holds. If the test is not statistically significant, it suggests that the unconfoundedness assumption may hold, making the causal inference problem likely solvable; in such cases, regression or machine learning methods can be applied to estimate causal effects. In contrast, a statistically significant result indicates a violation of the unconfoundedness assumption, making the causal effect unidentifiable.

There is limited literature on assessing the unconfoundedness assumption. Existing methods, developed in the 1980s \cite{Rosenbaum01061987}, used pseudo-treatments or pseudo-outcomes to evaluate its plausibility \cite{imbens2015causal}. A recent work used a Chi-square test \cite{yang2023elastic} to decide whether or not to combine real-world data with RCTs. To contribute a new method for assessing the unconfoundedness assumption, we propose a framework of hypothesis testing using two data sources: an observational dataset and an RCT dataset. We first formalize a mathematical distinction between two sets of parameters: those used in causal inference versus those used in statistical inference. To our knowledge, this distinction has not been explicitly formalized in the existing literature. More specifically, the two sets of parameters are:
\begin{itemize}
    \item One defined in terms of potential outcomes, which represents causal effects but is generally not estimable from observational data;
    \item Another defined in terms of conditional means, which is always estimable from data but may not reflect causal relationships.
\end{itemize}
We demonstrate that a fundamental issue of causal inference arises from the inequality between these two sets of parameters.

In the context of treatment effect estimation, violation of the unconfoundedness assumption is commonly attributed to the presence of unmeasured confounders. In this paper, we treat situations that satisfy the unconfoundedness assumption as equivalent to those with no unmeasured confounders (see Section \ref{sec:discussion}, Discussion, for further comments). We conduct the hypothesis testing by comparing two means, one from the RCT data and the other from the observational data. In the setting of a simple linear model for the treatment outcome, we identify several key factors that influence the power of the test. Interestingly, the test is not affected by the coefficient or magnitude of the unmeasured confounder but rather by its variance. This is an important and perhaps counterintuitive aspect of the problem. Our work relates to research on combining real-world data (RWD) with RCTs \cite{colnet2024causal,kallus2018removing,yang2020improved,yang2023elastic,wu2022integrative}, which is a promising strategy to accelerate drug discovery.

The rest of the paper is organized as follows. Section 2 introduces the distinction between causal and statistical parameters, explains the unconfoundedness assumption, and makes connection to causal structure graphs. Section 3 presents the proposed testing approach, establishes its control of Type I error and analyzes its power under linear models. Section 4 demonstrates the method through simulations and semi-synthetic data. Finally, Section 5 discusses the limitations of the hypothesis test and outlines directions for future work. The supplementary materials include the proofs of the lemma and theorem, the simulation settings, and the R code.

\section{Causal Inference for Treatment Effects}
\subsection{Notation and the fundamental issue of causal inference}
For each individual in the RCT or observational population, we consider a random tuple $(A,X,Y) \sim P$, where $X \in \mathcal{X} \subset \mathbb{R}^p$ denotes observed pre-treatment covariates, $Y \in \mathcal{Y}$ is a binary or continuous outcome, $A$ is the binary treatment assignment, with $A=0$ for the control and $A=1$ for the treatment individuals, and $P$ is from a distribution family defined on the corresponding sample space. In the context of treatment effect estimation, we are interested in the causal relationship between $A$ and $Y$. That is, a treatment effect is a causal effect, which captures more than mere statistical dependence.

There are two major approaches to causal inference: causal structure graphs and potential outcomes. We begin by describing the potential outcomes framework and defer the discussion of causal structure graphs to Section \ref{sec_graph}. The potential outcomes framework \cite{splawa1990application,rubin1977} represents one of the biggest contributions of the statistics community to causal inference. In particular, it provides a clear and formal definition of the causal effect of $A$ on $Y$. Specifically, for each individual, we define two potential outcomes, $Y(0)$ and $Y(1)$, representing the outcomes that would be observed under control and treatment, respectively. Importantly, only one of these two potential outcomes is observed for an individual, never both, as demonstrated below: 
\begin{equation}\label{form_Y}
Y=\left\{
\begin{aligned}
&Y(0), && \text{if } A=0,\\
&Y(1), && \text{if } A=1.
\end{aligned}
\right.
\end{equation}
In the causal inference context here, we are interested in the average treatment effect, defined as
\begin{equation}\label{form_tau}
    \tau = \mathbb{E}[Y(1)-Y(0)],
\end{equation}
or the conditional average treatment effect when we consider heterogeneity over $X$:
\begin{equation}\label{form_taux}
    \tau(x) = \mathbb{E}[Y(1)-Y(0)|X=x].
\end{equation}
We call them the average
causal effect (ACE) and the conditional average causal effect (CACE),
respectively, to emphasize that these are parameters specific to causal inference. However, since only one of the potential outcomes $Y(0)$ and $Y(1)$ is observed but never both, ACE or CACE are generally not estimable from data. 

The quantities that are estimable from the data \newline
$(A,X,Y) \sim P$ are a different set of parameters, defined as follows:
\begin{equation}\label{form_omega}
    \omega = \mathbb{E}[Y|A=1] - \mathbb{E}[Y|A=0],
\end{equation}
and
\begin{equation}\label{form_omegax}
    \omega(x) = \mathbb{E}[Y|A=1, X=x] - \mathbb{E}[Y|A=0, X=x].
\end{equation}
For instance, we can compute the difference in sample averages between the treatment and control groups and take it as an estimate of $\omega$. For an estimate of $\omega(x)$, regression methods or supervised machine learning approaches can be applied. 

\begin{table}[h]
    \caption{Causal inference versus statistical inference. These two types of inference involve distinct sets of parameters. While statistical parameters are estimable from data, the causal parameters are generally not identifiable from observational data.}
    \label{tab:parameters}
    \centering
    \begin{tabular}{|c|c|c|} \hline
         & Causal inference & Statistical inference  \\ \hline
   Parameters & $\tau$, $\tau(x)$ & $\omega$, $\omega(x)$ \\
   RCT & $\checkmark$ & $\checkmark$ \\
   Observational data & $\times$ & $\checkmark$ \\ \hline
    \end{tabular}
\end{table}

The fundamental issue of causal inference lies in the discrepancy between $\tau$
and $\omega$, or between
$\tau(x)$ and $\omega(x)$. From Formula (\ref{form_Y}), we have $Y=Y(0)\mathbf{1}_{A=0}+Y(1)\mathbf{1}_{A=1}$, where $\mathbf{1}_{(.)}$ is the indicator function. It is straightforward to verify that in order to obtain $\tau = \omega$, a sufficient assumption is $A \perp (Y(0), Y(1))$. Notably, this assumption is satisfied in an RCT, as the treatment assignment is random by design. This explains why an RCT is the gold standard to confirm a treatment effect, or causal effect in our context, because under an RCT the causal effect can be estimated through the estimation of $\omega$. We summarize the distinction between the two sets of parameters in Table \ref{tab:parameters}. They are essentially the difference between causal inference and traditional statistical inference.

\subsection{Unconfoundedness and existence of unmeasured 
confounders
}
The most common approach to estimating $\tau$ and $\tau(x)$ is through the estimation of $\omega$ and $\omega(x)$ under the assumption of unconfoundedness. As mentioned earlier, an RCT satisfies $ A \perp (Y(0), Y(1))$, when the treatment assignment is purely random. However, this assumption does not hold in observational studies. To estimate the causal parameters from observational studies, we usually assume that within homogeneous subpopulations, the treatment assignment is effectively random. Formally, we assume
\begin{equation}\label{unconf}
    A \perp (Y(0), Y(1)) \hspace{1mm}|  X.
\end{equation}
In addition, it is assumed that the treatment assignment probability is bounded away from zero and one, ensuring sufficient overlap in the covariate distributions between the treatment and control groups. This requirement, together with the conditional independent assumption (\ref{unconf}), is referred to as the unconfoundedness assumption \cite{rosenbaum1983central}. It is straightforward to verify that under the unconfoundedness assumption, we have
\[
\tau(x)=\omega(x).
\]
Therefore, unconfoundedness provides a favorable setting for causal effect estimation, as the causal parameters become identifiable and can be consistently estimated from observational data. Imbens \cite{imbens2024causal} provides a comprehensive review of methods for estimating $\tau$ \cite{chernozhukov2017double, rubin2006matched,robins2000marginal, hahn1998role} and $\tau(x)$ \cite{athey2016recursive, wager2018estimation} under unconfoundedness.

However, it is evident that the unconfoundedness assumption does not hold in many cases. Dropping this assumption means that causal effects become unidentifiable, and no general solution exists. One line of research addresses this challenge by relaxing the unconfoundedness assumption, allowing it to hold conditional on both observed and unmeasured covariates. Specifically, a modified unconfoundedness assumption is adopted: 
\begin{equation}\label{unmeasured}
 A \perp (Y(0), Y(1)) \hspace{1mm} | X, U   
\end{equation} 
where $U$ represents a set of unmeasured confounders. In Section 3, we propose a framework to test the existence of unmeasured confounders using two datasets: an RCT and a observational dataset. If the test is insignificant, it becomes reasonable to assume the unconfoundedness assumption holds. In that case, we estimate treatment effects through the estimation of $\omega(x)$ and $\omega$.

\subsection{Causal structure graphs}\label{sec_graph}
We now introduce the other major approach to causal inference: causal structure graphs, also referred to as structural causal models \cite{pearl2009causality}. These are directed acyclic graphs (DAGs), where nodes represent variables and directed edges encode cause-effect relationships. A causal structure graph defines a data generating process, as each node variable, except the root nodes (variables without parents), can be generated from a conditional distribution given its parent nodes. The joint distribution of all variables can be expressed as the product of these conditional distributions and the marginal distributions of root nodes

\begin{figure}[h]
  \centering
  \includegraphics[width=48mm]{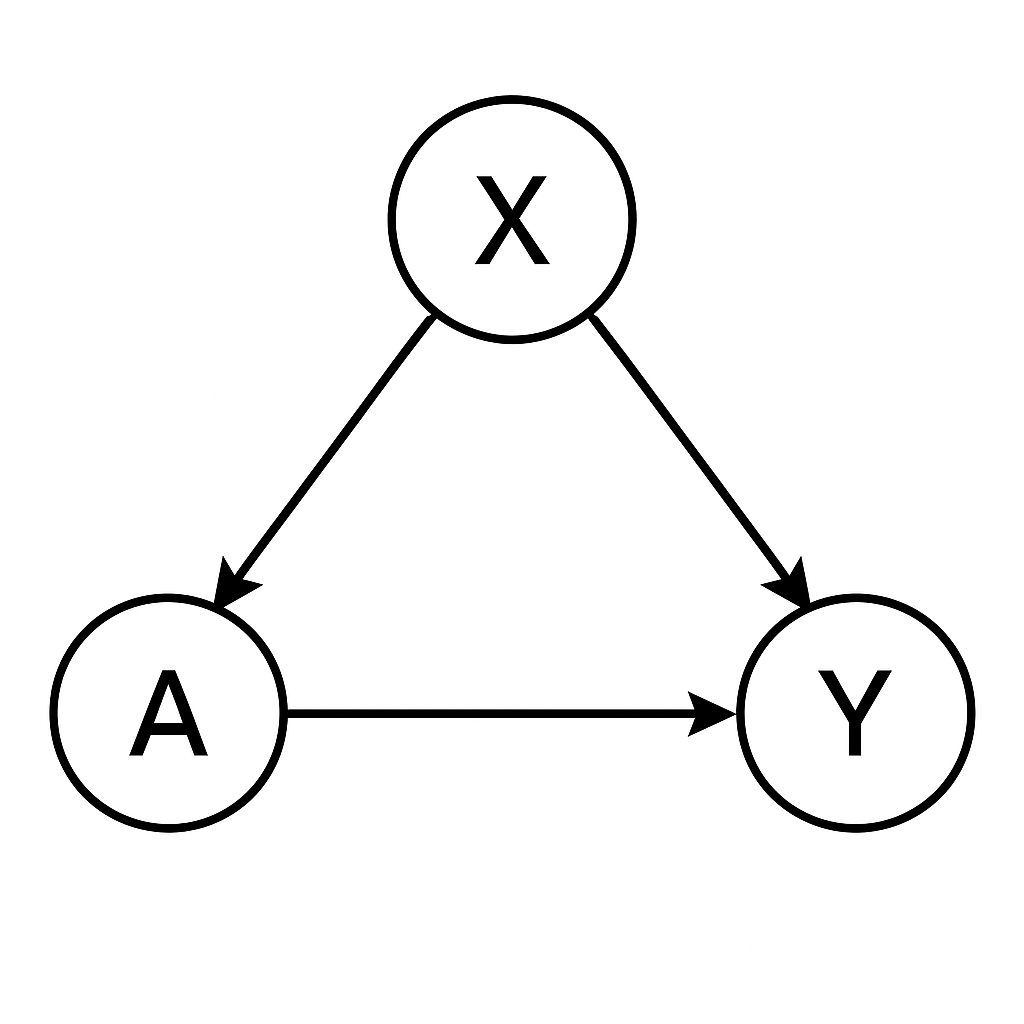}
  \caption{A causal structure graph illustrating cases corresponding to the unconfoundedness assumption.}
  \label{fig:causal-original}

  \vspace{0.5cm} 

  \includegraphics[width=57mm]{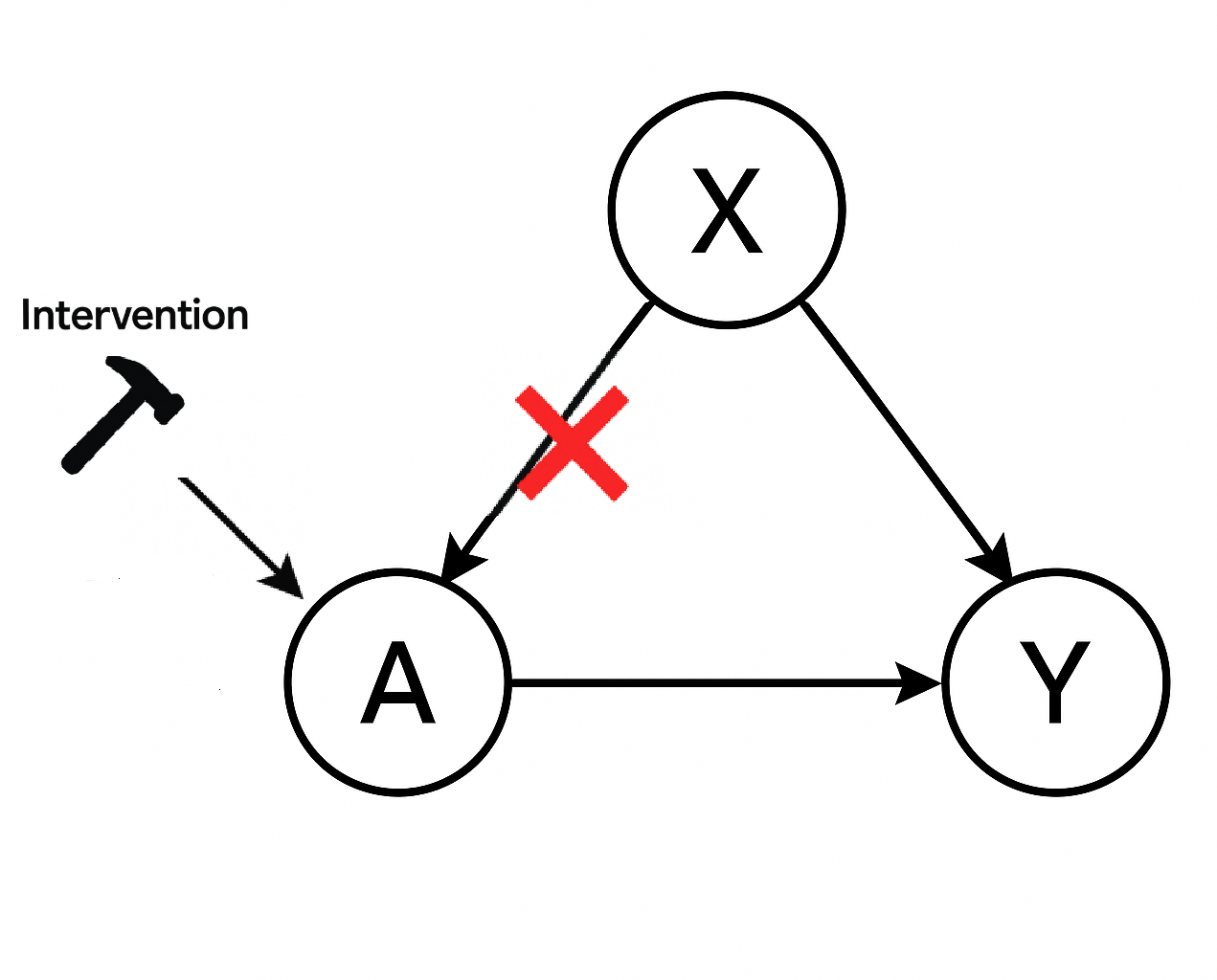}
  \caption{An RCT corresponds to an intervention on the treatment variable $A$.}
  \label{fig:causal-intervention}
  
\end{figure}

We illustrate two specific causal structure graphs for $(A, X, Y)$ in Figure \ref{fig:causal-original} and \ref{fig:causal-intervention}. In these graphs, the directionality from parents to children defines conditional distributions, which together with the marginal distributions of the root nodes determine the joint distribution $P(A, X, Y)$. Given such a DAG, we can specify a family of distributions for $P$. However, given data alone, we can only make statistical inferences about $P$, while the underlying DAG is generally not identifiable, because different DAGs can correspond to the same distribution $P$.

Causal structure graphs, 
rooted in graphical models, exploit the correspondence between graph separation and conditional independence to make causal inference. In a DAG, the absence of a direct path between two variables (nodes) corresponds to conditional independence given the separating variables (nodes). While causal effects cannot be directly identified from data alone, if conditional independence relationships can be reliably inferred from the data, causal inference becomes possible. More specifically, by conditioning on the separating nodes, we can test whether a pair of nodes are conditionally independent. If they are conditionally independent, any dependence between them is fully explained by the separating nodes. If they are conditionally dependent, the remaining dependence suggests a direct causal relationship between them. In other words,
by conditioning on appropriate covariates (nodes), one can block spurious paths, which are called back-door paths \cite{pearl2009causality}, leading to identification of the causal path of interest. 

For example, Figure \ref{fig:causal-original} illustrates a case where the causal effect from $A$ to $Y$ is identifiable. In this graph, $X$ represents a set of confounders, and there are no other unmeasured confounders. By conditioning on $X$, all back-door paths (confounding effects) from treatment $A$ to outcome $Y$ are blocked, allowing the direct causal path from $A$ to $Y$ to be identified. Given a dataset, the causal effect of $A$ on $Y$ can be identified by estimating the conditional means of $Y$ given $X$ and $A$. That is, either $A$ has no effect on $Y$, or it has a nonzero causal effect. 

Separately, the unconfoundedness assumption (\ref{unconf}), based on the potential outcomes framework, corresponds to the same causal structure shown in Figure \ref{fig:causal-original}. However, this specific causal structure is not hold in general, particularly when unmeasured confounders are present. 

Figure \ref{fig:causal-intervention} illustrates the RCT setting, where the random assignment of treatment or control is an intervention. This intervention breaks the dependence of $A$ on $X$, leaving the causal path from 
$A$ to $Y$ as the only remaining pathway. 

\section{Assessment of the unconfoundedness assumption}
\subsection{Testing framework with two datasets}
Direct assessment of the unconfoundedness assumption (\ref{unconf}) is impossible because we only observe $Y$, not the potential outcomes $Y(0)$ and $Y(1)$. We address this challenge by comparing the parameter $\omega$ (\ref{form_omega}) estimated from two datasets: one from an RCT and the other from an observational study. We first extend the notation introduced in Section 2.1 to the case of two datasets. More specifically, let $\mathcal{D}^r=\{(A_i, X_i, Y_i)\}_{i=1}^{n}$ be i.i.d. samples from the RCT with 
distribution $P^r$, where $A_i\in \{0,1\}$, $X_i \in \cal{X}$, and $Y_i \in
\cal{Y}$. Similarly, let 
$\mathcal{D}^o=\{(A_i, X_i, Y_i)\}_{i=n+1}^{n+m}$ be i.i.d. samples from the observational study, drawn from the same sample space but following a different
distribution $P^o$. For example, $P^o$ is defined by a data generating process from Figure \ref{fig:causal-original} and $P^r$ from Figure \ref{fig:causal-intervention}. It is evident that $P^r(A,X,Y) \ne P^o(A,X,Y)$. We assume that the RCT and observational datasets share the same sample space, and we comment on overlapping sample spaces in the Discussion section \ref{sec:discussion}. 

We further denote the estimable parameters, as defined in Formula (\ref{form_omega}) and (\ref{form_omegax}), for the two datasets as $\omega^e$ and $\omega^e(x)$,
where $e$ is denoted as either $r$ or $o$, referring to the RCT and the observational study, respectively. From the discussions in Section 2, we know that for  an RCT with pure randomized design,
\[
\omega^r = \tau.
\]
If the RCT satisfies the unconfoundedeness assumption (\ref{unconf}), for instance, with blocked randomized design, we have
\[
\omega^r(x) = \tau(x).
\]
On the other hand, for the observational data, we do not expect $\omega^o(x) = \tau(x)$ unless the unconfoundedness assumption (\ref{unconf}) holds. Here $\tau$ and $\tau(x)$ are the causal parameters, i.e., ACE and CACE. These causal parameters, defined through $(Y(0), Y(1))$, reflect the underlying causal relationship between $A$ and $Y$ and remain the same across the two datasets, despite differences between $P^r$  and $P^o$ . 

Now, we can assess the unconfoundedness assumption by testing whether $\omega^r(x) = \omega^o(x)$, or more simply, whether $\omega^r = \omega^o$ after averaging over $x \in \cal{X}$. Given the two datasets, we first obtain estimators $\hat{\omega}^r$ and $\hat{\omega}^o$, and then test the following hypotheses, for example, via asymptotic normal tests or bootstrap tests: 
\begin{align*}
    H_0 &: \omega^o=\omega^r \\
    H_a &: \omega^o\ne\omega^r.
\end{align*}
If the null hypothesis cannot be rejected based on the two datasets, it suggests that $\omega^o = \omega^r = \tau$, and it is plausible that $\omega^o(x) = \omega^r(x) = \tau(x)$ for $x \in \mathcal{X}$. This indicates that the unconfoundedness assumption holds and there are no unmeasured confounders in the observational dataset. Therefore, the observational data can be combined with the RCT data to improve causal effect estimation.

\bigskip

\begin{algorithm}
  \caption{Bootstrap Test for Unconfoundedness Assumption}
  \label{alg:bootstrap}
  \begin{algorithmic}[1]
    \REQUIRE $\mathcal{D}^r,\mathcal{D}^o,B,\text{Significant Level  } \alpha$
    \FOR{$b=1$ \TO $B$}
      \STATE Generate $\mathcal{D}^r_b$ and $\mathcal{D}^o_b$ by resampling with replacement.
      \STATE $\hat{\omega}^r_b \gets \text{Estimator of}$ $ \omega^r(\mathcal{D}^r_b)$, $\hat{\omega}^o_b \gets \text{Estimator of}$ $ \omega^o (\mathcal{D}^o_b)$
      \STATE $T_b \gets \hat{\omega}^o_b - \hat{\omega}^r_b$
    \ENDFOR
    \STATE $\hat{q}_{\alpha/2}\gets\text{Empirical Quantile}(\{T_b\},\alpha/2)$,\ 
    
          $\hat{q}_{1-\alpha/2}\gets\text{Empirical Quantile}(\{T_b\},1-\alpha/2)$
    \STATE \textbf{if }$ 0 \notin [\hat{q}_{1-\alpha/2}, \hat{q}_{\alpha/2}]$\ \textbf{then} \text{Reject} $H_0$\ \textbf{else} Fail to reject $H_0$.
  \end{algorithmic}
\end{algorithm}

We present the bootstrap procedure in Algorithm \ref{alg:bootstrap}. Denote the sample size of the RCT data $\mathcal{D}^r$ by $n$ and that of the observational data $\mathcal{D}^o$ by $m > n$. Let $\hat{\omega}^o$ and $\hat{\omega}^r$ be consistent estimators of $\omega^o$ and $\omega^r$, respectively. Define the test statistic as 
$$T_{n,m} = \hat{\omega}^o - \hat{\omega}^r.$$
Under $H_0$,  $T_{n,m}$ converges to zero. Denote its cumulative distribution function (CDF) by
$$F_{n,m}(t) = P(T_{n,m}<t | H_0).$$ 
The bootstrap procedure resamples separately within $\mathcal{D}^r$ and $\mathcal{D}^o$ and computes
$$T_{b}^* = \hat{\omega}_b^o-\hat{\omega}_b^r, \hspace{5mm} b=1,\dots,B. $$
Denote the bootstrap CDF
$$F_{n,m}^*(t) = P(T_{b}^*<t|\mathcal{D}^r,\mathcal{D}^o).$$ 
By the bootstrap consistency result \cite{ctx36517209530001081,Romano_2012,tibshirani1993introduction}, we have
$$\sup_t|F_{n,m}^*(t)-F_{n,m}(t)| \overset{p}\longrightarrow 0 \quad (n,m\rightarrow\infty).$$
Since we only use a finite number $B$ in the procedure, instead of knowing the bootstrap CDF $F_{n,m}^*(t)$, we compute $T^*_1,...,T^*_B$ and obtain the empirical bootstrap CDF
$$\hat{F}_B(t) = \frac{1}{B}\sum_{b=1}^B1\{T^*_b < t\}.$$
The quantiles of $F^*_{n,m}$ are estimated by the empirical quantiles $\hat{q}_{1-\alpha/2}$ and $\hat{q}_{\alpha/2}$ using $\hat{F}_B(t).$ 
It has been proven in \cite{tibshirani1993introduction} that $$P(T_{n,m}>\hat{q}_{1-\alpha/2})+P(T_{n,m}<\hat{q}_{\alpha/2}) \longrightarrow \alpha +\mathcal{O}(B^{-1/2}),$$
when $n,m\rightarrow\infty$. In practice, choosing $B^{-1/2} \ll n^{-1/2}$ makes this deviation negligible. In other words, it ensures that the Type I error of the bootstrap test can be controlled at $\alpha$ asymptotically. 

\subsection{Power analysis under a linear model}
\begin{figure}[h]
  \centering

    \centering
    \includegraphics[width=45mm]{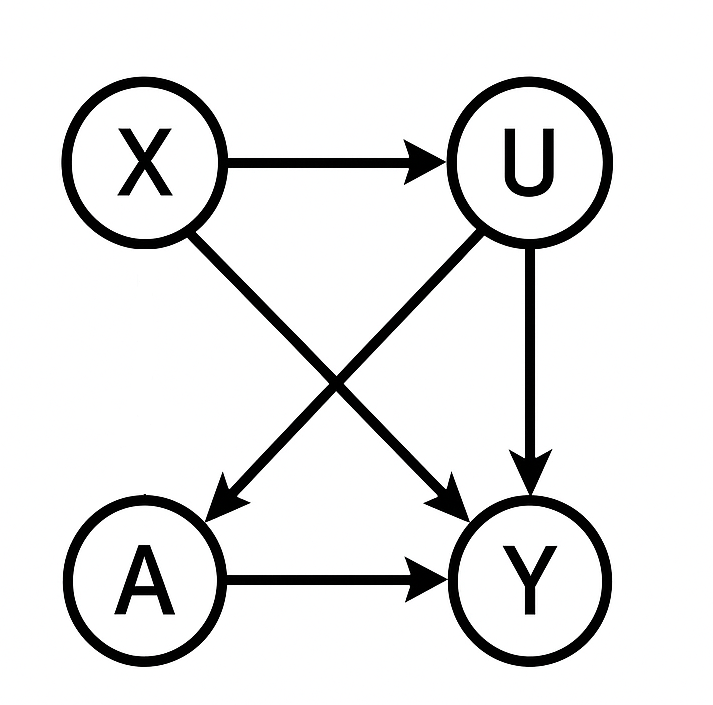}
    \caption{DAG of the observational study with unmeasured confounder $U$.}
    \label{fig:obs}

    \centering
    \includegraphics[width=40mm]{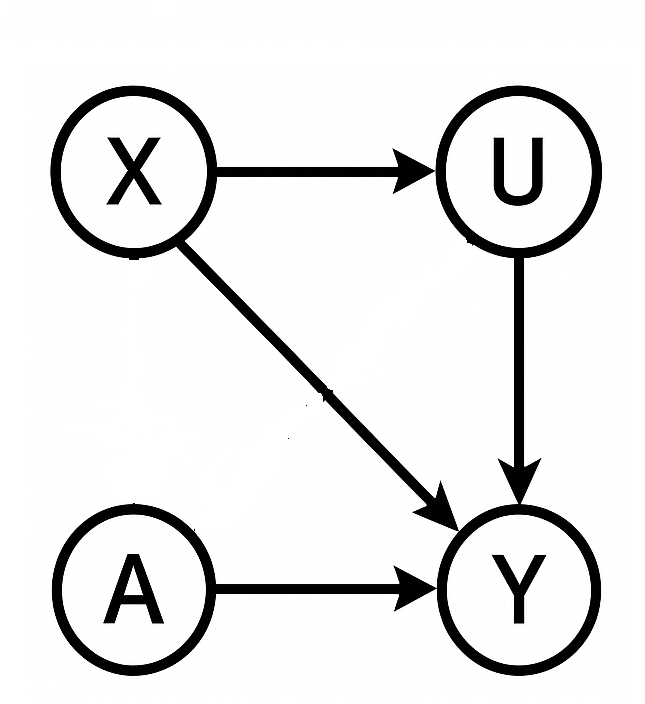}
    \caption{DAG of the RCT.}
    \label{fig:rct}
  \label{fig:linear study}
\end{figure}

To gain further insight into the performance of our testing procedure, we analyze its power under linear models. While power analysis is generally challenging, the linear setting provides a tractable case for theoretical derivation. Suppose there is an observational study with an underlying model:
\[
Y = \beta_0 + \beta_A A + \beta_X X + \beta_U U + \epsilon,\]
\[\epsilon\sim \mathcal{N}(0,\sigma_\epsilon^2),\quad X \sim \mathcal{N}(\mu_X,\sigma_X^2). \nonumber
\]
 The true average causal effect (ACE) here is $\tau=\beta_A$. However, in the observational study, we do not observe the unmeasured confounder $U$, we actually estimate $\beta^{'}_A$ in the model
\begin{equation}\label{estimable}
Y = \beta^{'}_0 + \beta^{'}_A A + \beta^{'}_X X + \zeta,
\end{equation}
where $\zeta$ denotes the error term. Suppose the unobserved confounder has an underlying model:
\[
    U = \delta_0 + \delta_A A + \delta_X X + \nu, \quad \nu\sim \mathcal{N}(0,\sigma_U^2).
\]
The observational data (OBS) can be used to fit a model as (\ref{estimable}), thus the estimable parameter is $\omega^o = \beta^{'}_A=\beta_A + \beta_U\delta_A$. Figure \ref{fig:obs} presents the causal graph, or DAG, of the observational study. 
In an RCT shown in Figure \ref{fig:rct}, the distribution of $(X, U)$ remains the same as OBS, but the treatment $A$ is randomly assigned and independent of $U$ and $X$. Therefore, observing $U$ or not does not affect the estimation of $\beta_A$  in the RCT, and $\omega^r = \tau = \beta_A$. 
The following formulas represent two different parameters from the two datasets:
\begin{equation}\label{bias}
\left\{
\begin{aligned}
\omega^o &= \beta_A + \beta_U\,\delta_A && \text{from OBS},\\
\omega^r &= \tau = \beta_A               && \text{from RCT}.
\end{aligned}
\right.
\end{equation}
It is clear that, in the observational study, the estimable parameter $\omega^o$ can be decomposed into two terms: the true ACE $\beta_A$ and the bias (or confounding effect) $\beta_U  \delta_A$ induced by the unmeasured confounder.

We use the inverse probability weighted (IPW) estimator to estimate $\omega^r$ and $\omega^o$ in the RCT and observational study:  
\begin{equation}
\label{eq:ipw}
    \hat{\omega} =\frac{\sum_i^{n} A_iY_i/e(X_i)}{\sum_i^{n} A_i/e(X_i)} - \frac{\sum_i^{n} (1-A_i)Y_i/(1-e(X_i))}{\sum_i^{n} (1-A_i)/(1-e(X_i))},
\end{equation}
  where $e(X)$ is the propensity score \cite{rosenbaum1983central}. 
 
Additionally, we denote (1) $p_o = P^o(A=1)$ in the observational study and $p_r = P^r(A=1)$ in the RCT; (2) $c=\beta_X/\beta_U$ to be some finite constant, which means $X$ and $U$ are driving the outcome in the same scale; (3) the noise ratio $\eta = \sigma^2_{\epsilon}/\beta^2_U$ to be a small number. In the study described above, we have the following Lemma \ref{Lemma:linear} about the IPW estimators $\hat{\omega}^r$ and $\hat{\omega}^o$.
\begin{lemma}
\label{Lemma:linear}
The inverse probability weighted (IPW) estimator
for the RCT and observational data have the following asymptotic distributions:
$$\sqrt{m}(\hat{\omega}^o -\beta_A-\beta_U\delta_A)\overset{d}\longrightarrow \mathcal{N}(0, V_o)$$
where $V_o/m = \mathrm{Var}(\hat\omega^o)$. Analogously,   
$$\sqrt{n}(\hat{\omega}^r -\beta_A)\overset{d}\longrightarrow \mathcal{N}(0, V_r)$$
where $V_r/n = \mathrm{Var}(\hat\omega^r)$. 
\end{lemma}
 Lemma \ref{Lemma:linear} implies that under $H_0$ of $\omega^o = \omega^r$, we have
$$\frac{\hat{\omega}^o-\hat{\omega}^r}{\sqrt{\frac{V_o}{m}+\frac{V_r}{n}}}\overset{d}\longrightarrow \mathcal{N}(0, 1).$$
By plugging in consistent estimators of $\hat{V}_o$ and $\hat{V}_r$, a $Z$ statistics as defined below has an asymptotically standard normal distribution under $H_0$:
$$Z = \frac{\hat{\omega}^o-\hat{\omega}^r}{\sqrt{\frac{\hat{V}_o}{m}+\frac{\hat{V}_r}{n}}}.$$
Thus we can perform a $z$-test at level $\alpha$ by rejecting $H_0$ when $|Z| > z_{1-\frac{\alpha}{2}}$. This $z$-test is able to assess the unconfoundedness assumption. Moreover, for the asymptotically $z$-test, we can also derive its power analytically. 

Different methods for estimating the propensity score in the IPW estimator will affect $V_r$ and $V_o$, and thus the form of the power function. For example,
\begin{itemize}
    \item for the RCT, the propensity score in $\hat\omega^r$ is estimated by a constant, i.e., $e(X_i) = \hat{p}_r$.
    \item for the OBS, the propensity score in $\hat\omega^o$ is estimated by logistic regression, i.e., $e(X_i)=\mathrm{expit}(\gamma_0+\gamma_1 X_i) = 1/\left(1+\mathrm{e^{-(\gamma_0+\gamma_1 x_i)}}\right)$.
\end{itemize}
Theorem \ref{Theorem:LinearPower} characterizes the power function when the propensity scores are estimated using the above methods.

\begin{theorem}
\label{Theorem:LinearPower}
 We can perform a $z$-test at level $\alpha$ by rejecting $H_0$ when $|Z| > z_{1-\frac{\alpha}{2}}$. The power function of the test will have the form:
$$ \Phi\left(-z_{1-\frac{\alpha}{2}}- \frac{\delta_A}{h}\right) + \Phi\left(\frac{\delta_A}{h} -z_{1-\frac{\alpha}{2}}\right) + \mathcal{O}(n^{-\frac{1}{2}}),$$
where $\Phi$ is the CDF of the standard normal distribution and $h=h(\delta_A,\delta_X,\sigma_v,\sigma_{\epsilon},n)$ is defined as follows with $\kappa = \frac{m}{n}$.

\begin{equation*}
\begin{aligned}
h(\delta_A,\delta_X,\sigma_v,\sigma_{\epsilon},n) = \left\{\frac{(c+\delta_X)^2\sigma_X^2 + \delta_A^2\,p_o(1-p_o)+\sigma_U^2+\eta}{np_r(1-p_r)\,}\right.\left.+  \frac{\sigma_U^2+\eta}{\kappa np_o(1-p_o)}\Bigr]
\right\}^{1/2}.
\end{aligned}
\end{equation*}
\end{theorem}

\begin{remark}
We identify several key factors that influence the power of the test by Theorem \ref{Theorem:LinearPower}. The theorem implies that the magnitude of $\beta_U$ (the coefficient of the unmeasured confounders) does not influence the power given $c=\beta_X/\beta_U$ and $\eta = \sigma^2_{\epsilon}/\beta^2_U$. 
\end{remark}

The key factors are $n$ (the sample size of the RCT data), $\delta_A$ and $\delta_X$ (the effect sizes of $A$ and $X$ on the unmeasured confounder), $\sigma^2_U$ and $\sigma^2_X$ (the variances). Specifically, when $\kappa \rightarrow \infty$, we have:
\begin{equation*}
\begin{aligned}
h(\delta_A,\delta_X,\sigma_v,\sigma_{\epsilon},n)=\sqrt{\frac{(c + \delta_X)^2\sigma_X^2 +\delta_A^2p_o(1-p_o) +\sigma_U^2+ \eta}
     {n\,p_r(1-p_r)}
   }.
\end{aligned}
\end{equation*}
   Given the form of $h$, increasing $n$ and $\delta_A$ will cause the power to increase while increasing $\delta_X$ and variance of the confounders will cause the power to decrease. Proofs of Lemma \ref{Lemma:linear} and Theorem \ref{Theorem:LinearPower} are included in the supplementary materials.

\section{Experiments}
\subsection{Simulated Data}

In this subsection, we evaluate how well our bootstrap procedure controls the Type I error rate using synthetic datasets generated from both a linear model and a non-linear model.

\textbf{Linear model}
To simulate observational study data, $m$ samples will be generated. We first generate $X$ and $A$ by
\[
X \sim \mathcal{N}(\mu_X,\sigma_X^2), \quad A \sim \mathrm{Bernoulli}(p_o),
\]
and then generate the unobserved confounder via
\[
U = \delta_0 + \delta_A A + \delta_X X + \nu, \quad \nu\sim \mathcal{N}(0,\sigma_U^2).
\]
The last step is to generate outcome as follows
\begin{align}
\label{eq:outcome_generate}
\begin{aligned}
    &Y = \beta_0 + \beta_A A + \beta_X X + \beta_U U + \epsilon, \\
    &\epsilon\sim \mathcal{N}(0,\sigma_\epsilon^2).
    \end{aligned}
\end{align}
This generative process results in an observational dataset with a structure consistent with the causal graph in Figure \ref{fig:obs}.
For the RCT data with sample size $n$, $X$, $A$ and $U$ are generated by the same distributions as in the observational study first. However, after $X$ and $U$ are obtained, the treatment $A$ is randomly replaced according to another distribution $A \sim \mathrm{Bernoulli}(p_r)$, ensuring independence between  $U$ and $A$. The outcome is still generated by Formula (\ref{eq:outcome_generate}). The generated RCT dataset has a causal graph as Figure \ref{fig:rct}.
 
Under the data-generating model, the null hypothesis $H_0:\omega^o = \omega^r$ holds when $\beta_U\delta_A=0$. We examine two scenarios that satisfy the null hypothesis: $\beta_U = 0$ or $\delta_A=0$. For each scenario, we generate 
$1,000$ independent replicates; in each replicate we generate an RCT sample of size $n=100$ and an observational sample of size $m = 2,000$. We then apply the proposed bootstrap test with $B=1,000$ resamples and compute the empirical Type I error as the proportion of replicates in which $H_0$ is rejected at a significant level $\alpha$. Detailed parameter settings are reported in the supplementary material.

\textbf{Non-linear model}
The non-linear model has a similar data generating procedure with the linear case except that $U$ and $Y$ have different underlying models as follows 
    \[
    U = \delta_0 + \delta_X X^2 + \delta_A A + \delta_{XA}XA + \nu
    \]
     \[Y = f(X,A) + \beta_Ug(U) + \epsilon,
    \]
    \[
     f(X,A) = \gamma_0 + \gamma_X X^2 + \gamma_A A + \gamma_{XA} X^2A,
     \quad
    g(U) = U^2.
    \]
Under this data-generating model, the null hypothesis $H_0:\omega^o = \omega^r$ holds when: $(\beta_U,\delta_A,\delta_{XA})=(0,\delta_A,\delta_{XA})$ or $(\beta_U,\delta_A,\delta_{XA})=(\beta_U,0,0)$. Again, we examine two scenarios by generating $1,000$ independent replicates for each scenario; in each replicate we generate an RCT sample of size $n=100$ and an observational data sample of size $m = 2,000$. We then apply the proposed bootstrap test with $B=1,000$ resamples and compute the empirical Type I error as the proportion of replicates in which $H_0$ is rejected at significant level $\alpha$. Detailed parameter settings are reported in the supplementary material.

Tables \ref{tab:type1_linear} and \ref{tab:type1_nonlinear} report the empirical Type I error rates (i.e.\ proportion of rejections) for the linear and non-linear model at several significant levels $\alpha$. The results demonstrate that our test controls the Type I error at the corresponding level $\alpha$.

\begin{table}[h]
\caption{Type I Error of Bootstrap for the Linear Model at Different $\alpha$ }
\label{tab:type1_linear}
\vskip5mm\centering
\setlength{\tabcolsep}{4pt} 
\begin{tabular}{ccc}\hline
 & $(\beta_U,\delta_A) = (2,0)$ & $(\beta_U,\delta_A) = (0,2)$ \\\hline
 $\alpha=0.1$ & $0.099$ & $0.101$  \\\hline
 $\alpha=0.05$ & $0.051$ & $0.053$   \\\hline
 $\alpha=0.01$ & $0.012$ & $0.010$   \\\hline
\end{tabular}
\end{table}
\begin{table}[H]
\caption{Type I Error of Bootstrap for the Non-linear Model at Different $\alpha$ }
\label{tab:type1_nonlinear}
\vskip5mm\centering
\setlength{\tabcolsep}{4pt} 
\begin{tabular}{ccc}\hline
 & $(\beta_U,\delta_A,\delta_{XA})=(0,2,2)$ & $(\beta_U,\delta_A,\delta_{XA})=(2,0,0)$ \\\hline
 $\alpha=0.1$ & $0.102$ & $0.099$  \\\hline
 $\alpha=0.05$ & $0.051$ & $0.050$   \\\hline
 $\alpha=0.01$ & $0.009$ & $0.011$   \\\hline
\end{tabular}
\end{table}

\begin{figure}[h]
  \centering
    \includegraphics[width=80mm]{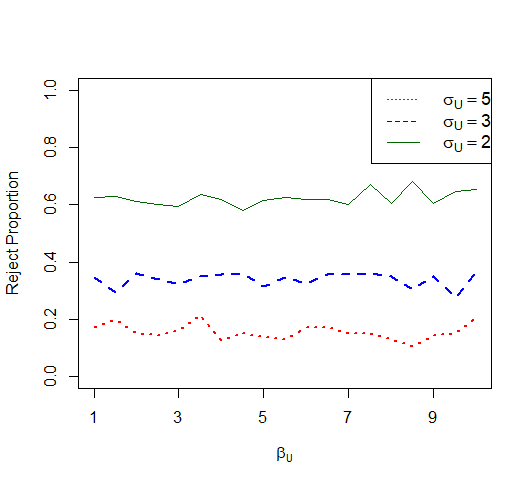}
    \caption{Power (defined by reject proportion) vs.\ \(\beta_U\) (the coefficient of the unmeasured confounder on $Y$). Given $\delta_A = 1$, $\omega^o - \omega^r =\beta_U$ under $H_a$. This figure indicates that $\beta_U$ has no impact on power.}
    \label{fig:power_beta_u}
  \hfill

    \includegraphics[width=80mm]{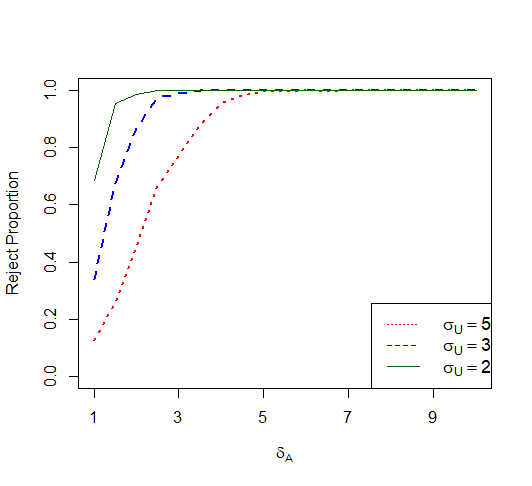}
    \caption{Power (defined by reject proportion) vs.\ \(\delta_A\) (the coefficient of treatment assignment on the unmeasured confounder $U$). Given $\beta_U = 1$, $\omega^o - \omega^r =\delta_A$ under $H_a$. This figure indicates that $\delta_A$ has a positive impact on power.}
    \label{fig:power_delta_a}

\end{figure}

To demonstrate how $\beta_U$ and $\delta_A$ affect the power of the test and validate Theorem \ref{Theorem:LinearPower}, we generate both RCT data and observational data with an unmeasured confounder using the linear model described above under the alternative hypothesis $H_a$. Since $\omega^o - \omega^r = \beta_U\delta_A$, we vary either $\beta_U$ (Figure \ref{fig:power_beta_u}) or $\delta_A$ (Figure \ref{fig:power_delta_a}) while holding the other fixed. At the significant level $\alpha=0.05$, empirical power using 200 replicates for each value of $\beta_U$ or $\delta_A$ with $B=500$ bootstrap resamples per replicate. Detailed parameter settings are reported in the supplementary materials. Fig \ref{fig:power_beta_u} demonstrates our conclusion that $\beta_U$ has no impact on the power and Fig \ref{fig:power_delta_a} demonstrates that $\delta_A$ has a positive impact on power.

\subsection{Real World Data}

The Tennessee Student/Teacher Achievement Ratio (STAR) project \cite{word1990state} was a large-scale randomized controlled trial (RCT) conducted in the late 1980s to assess the effect of class size on student academic performance, as measured by standardized test scores (SAT). We obtain the STAR project dataset from an example in \textit{Applied Econometrics with R} \cite{kleiber2008applied}. In the STAR project, both students and teachers were randomly assigned to one of three classroom settings: small classes (13–17 students), regular-sized classes (22–25 students), and regular-sized classes with a full-time teacher's aide. These assignments, initiated in kindergarten, were maintained through the third grade. The results demonstrated that students in smaller classes experienced significant improvements in standardized test scores, particularly in reading and mathematics \cite{krueger1999experimental, word1990state}.

Since many students entered the study beginning in the first grade, we define treatment based on their class type in the first grade. For the purposes of our analysis, we consider a binary treatment variable: assignment to a small class ($A=1$) versus a regular class ($A=0$). The regular-sized classes with a full-time teacher's aide are not included in our study. The outcome variable $Y$ is defined as the sum of reading and math SAT scores at the end of the first semester. The covariates $X$ include gender, race, birth year, free lunch given or not, school location. After excluding observations with missing values, a total of $4,165$ samples are included in our analysis and $1,498$ of them are assigned to the treatment group.

In our analysis, we utilize both an RCT and an observational dataset. The RCT is formed by simple random sampling of 200 samples from the original dataset. From the remaining subjects, we construct the observational dataset and induce confounding using the school-location covariate (rural, inner-city, urban, suburban), which is known to affect outcomes \cite{kallus2018removing, krueger1999experimental}. Specifically, we apply the following selection rule: In ``urban'' or ``suburban'' schools, we include all control subjects ($A = 0$) but only high-preforming treated subjects ($A = 1$) whose outcomes $Y$ fall in the upper half among treated subjects in these schools; In ``rural'' or ``inner-city'' schools, we include all treated subjects ($A = 1$) but only control subjects ($A = 0$) whose outcomes $Y$ fall in the lower half among control subjects in these schools. This procedure induces confounding by making school location a confounder that influences both treatment assignment and outcomes in the observational study. By excluding school location from the dataset, we create  unmeasured confounding in the observational study.

\begin{figure}[H]
    \centering
    \includegraphics[width=80mm]{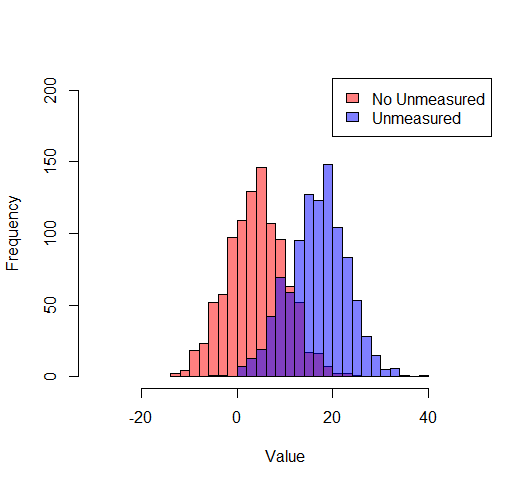}
  \caption{The empirical distribution of $\hat{\omega}^o - \hat{\omega}^r$: comparison with and without an unmeasured confounder. When an unmeasured confounder is present, the $95\%$ bootstrap confidence interval does not include 0, allowing us to conclude that $\omega^o - \omega^r$ is significantly different from 0.}
  \label{fig:rwd_hist}
\end{figure}

To estimate $\omega^r$ and $\omega^o$, we use the IPW estimator as Formula (\ref{eq:ipw}). Logistic regression is applied to estimate the propensity score. To evaluate our test, we compare results under two scenarios: (1) school location is observed and included as a covariate (i.e., no unmeasured confounder), and (2) school location is unobserved (i.e., presence of unmeasured confounder). When school location is observed, we obtain that $\hat{\omega}^r=25.38$ and $\hat{\omega}^o=29.64$; while when school location is unobserved, we obtain that $\hat{\omega}^r=24.65$ and $\hat{\omega}^o=41.39$. We generate the empirical distribution of $\hat{\omega}^o - \hat{\omega}^r$ by using the bootstrap procedure and show it in Figure \ref{fig:rwd_hist}. When there is no unmeasured confounder, $\hat{\omega}^o - \hat{\omega}^r$ does not deviate from 0 significantly. However, when an unmeasured confounder is present, the $95\%$ bootstrap confidence interval does not include 0, allowing us to conclude that $\omega^o - \omega^r$ is significantly different from 0. This also demonstrates the effectiveness of our bootstrap procedure on detecting unmeasured confounders on real-world data.

\section{Discussion}\label{sec:discussion}
The method developed in this work provides a practical approach to evaluating the unconfoundedness assumption. If the test result is not statistically significant, it suggests that the assumption may hold, making the causal inference problem likely solvable. In such cases, statistics methods, such as the IPW estimator, or machine learning approaches can be applied to estimate causal effects from observational data. Conversely, a statistically significant result indicates a violation of the assumption, implying that the causal effect is not identifiable from the observational data.

When the null hypothesis is true, it is likely that the unconfoundedness assumption is satisfied. However, multiple scenarios fall under the alternative hypothesis. These include the presence of unmeasured confounders as well as the inclusion of bad controls \cite{Cinelli2020}. As stated in \cite{imbens2024causal}, ``In practice, using variables causally affected by the treatment or outcome is the most common mistake in choosing variables to condition on in estimating average treatment effects using unconfoundedness approaches''. The proposed test is valid for testing the null hypothesis, regardless of the alternative hypothesis scenarios. Therefore, it extends beyond testing for unmeasured confounders.

This research work connects with machine learning problems of multiple domains, particularly in relation to out-of-distribution (OOD) generalization. The current method requires that the RCT and the observational study share the same sample space, a condition that is also common in most OOD generalization approaches. We are developing methods to handle situations where the sample spaces of the RCT and observational data differ.

When the proposed test rejects $H_0$, one possible direction is to apply a deconfounding method to adjust for the confounding bias. For example, the bias estimation approach described in \cite{kallus2018removing} could be employed as a procedure to improve causal effect estimation. This falls within the scope of our other research work.

\section*{Supplementary Material}
Supplementary materials are available at \url{https://github.com/muyeliu1991/When-Causal-supplementary}.  The PDF file contains the proofs of Lemma \ref{Lemma:linear} and Theorem \ref{Theorem:LinearPower}, as well as tables of parameter settings for simulations. R files include datasets and R code to reproduce the simulation results in Section 4.

\bibliographystyle{unsrt}  

\bibliography{reference}

\begin{thebibliography}{10}

\bibitem{pearl2009causality}
Judea Pearl.
\newblock {\em Causality}.
\newblock Cambridge university press, 2009.

\bibitem{spirtes2000causation}
Peter Spirtes, Clark~N Glymour, and Richard Scheines.
\newblock {\em Causation, prediction, and search}.
\newblock MIT press, 2000.

\bibitem{li2025featurematchinginterventionleveraging}
Haoze Li and Jun Xie.
\newblock Feature matching intervention: Leveraging observational data for causal representation learning.
\newblock {\em arXiv preprint arXiv:2503.03634}, 2025.

\bibitem{splawa1990application}
Jerzy Splawa-Neyman, Dorota~M Dabrowska, and Terrence~P Speed.
\newblock On the application of probability theory to agricultural experiments. essay on principles. section 9.
\newblock {\em Statistical Science}, pages 465--472, 1990.

\bibitem{rubin1977}
Donald~B. Rubin.
\newblock Assignment to treatment group on the basis of a covariate.
\newblock {\em Journal of Educational Statistics}, 2(1):1--26, 1977.

\bibitem{fisher1937design}
Ronald~Aylmer Fisher.
\newblock The design of experiments.
\newblock 1937.

\bibitem{scholkopf2021toward}
Bernhard Sch{\"o}lkopf, Francesco Locatello, Stefan Bauer, Nan~Rosemary Ke, Nal Kalchbrenner, Anirudh Goyal, and Yoshua Bengio.
\newblock Toward causal representation learning.
\newblock {\em Proceedings of the IEEE}, 109(5):612--634, 2021.

\bibitem{rosenbaum1983central}
Paul~R Rosenbaum and Donald~B Rubin.
\newblock The central role of the propensity score in observational studies for causal effects.
\newblock {\em Biometrika}, 70(1):41--55, 1983.

\bibitem{angrist1990lifetime}
Joshua~D Angrist.
\newblock Lifetime earnings and the vietnam era draft lottery: evidence from social security administrative records.
\newblock {\em The american economic review}, pages 313--336, 1990.

\bibitem{angrist1996identification}
Joshua~D Angrist, Guido~W Imbens, and Donald~B Rubin.
\newblock Identification of causal effects using instrumental variables.
\newblock {\em Journal of the American statistical Association}, 91(434):444--455, 1996.

\bibitem{imbens2014instrumental}
Guido Imbens.
\newblock Instrumental variables: an econometrician's perspective.
\newblock Technical report, National Bureau of Economic Research, 2014.

\bibitem{andrews2019weak}
Isaiah Andrews, James~H Stock, and Liyang Sun.
\newblock Weak instruments in instrumental variables regression: Theory and practice.
\newblock {\em Annual Review of Economics}, 11(1):727--753, 2019.

\bibitem{manski1990nonparametric}
Charles~F Manski.
\newblock Nonparametric bounds on treatment effects.
\newblock {\em The American Economic Review}, 80(2):319--323, 1990.

\bibitem{rosenbaum1983assessing}
Paul~R Rosenbaum and Donald~B Rubin.
\newblock Assessing sensitivity to an unobserved binary covariate in an observational study with binary outcome.
\newblock {\em Journal of the Royal Statistical Society: Series B (Methodological)}, 45(2):212--218, 1983.

\bibitem{rosenbaum2002observational}
Paul~R Rosenbaum.
\newblock Observational studies.
\newblock In {\em Observational studies}, pages 1--17. Springer, 2002.

\bibitem{Rosenbaum01061987}
Paul~R. Rosenbaum.
\newblock Model-based direct adjustment.
\newblock {\em Journal of the American Statistical Association}, 82(398):387--394, 1987.

\bibitem{imbens2015causal}
Guido~W Imbens and Donald~B Rubin.
\newblock {\em Causal inference in statistics, social, and biomedical sciences}.
\newblock Cambridge university press, 2015.

\bibitem{yang2023elastic}
Shu Yang, Chenyin Gao, Donglin Zeng, and Xiaofei Wang.
\newblock Elastic integrative analysis of randomised trial and real-world data for treatment heterogeneity estimation.
\newblock {\em Journal of the Royal Statistical Society Series B: Statistical Methodology}, 85(3):575--596, 2023.

\bibitem{colnet2024causal}
B{\'e}n{\'e}dicte Colnet, Imke Mayer, Guanhua Chen, Awa Dieng, Ruohong Li, Ga{\"e}l Varoquaux, Jean-Philippe Vert, Julie Josse, and Shu Yang.
\newblock Causal inference methods for combining randomized trials and observational studies: a review.
\newblock {\em Statistical science}, 39(1):165--191, 2024.

\bibitem{kallus2018removing}
Nathan Kallus, Aahlad~Manas Puli, and Uri Shalit.
\newblock Removing hidden confounding by experimental grounding.
\newblock {\em Advances in neural information processing systems}, 31, 2018.

\bibitem{yang2020improved}
Shu Yang, Donglin Zeng, and Xiaofei Wang.
\newblock Improved inference for heterogeneous treatment effects using real-world data subject to hidden confounding.
\newblock {\em arXiv preprint arXiv:2007.12922}, 2020.

\bibitem{wu2022integrative}
Lili Wu and Shu Yang.
\newblock Integrative $ r $-learner of heterogeneous treatment effects combining experimental and observational studies.
\newblock In {\em Conference on Causal Learning and Reasoning}, pages 904--926. PMLR, 2022.

\bibitem{imbens2024causal}
Guido~W Imbens.
\newblock Causal inference in the social sciences.
\newblock {\em Annual Review of Statistics and Its Application}, 11, 2024.

\bibitem{chernozhukov2017double}
Victor Chernozhukov, Denis Chetverikov, Mert Demirer, Esther Duflo, Christian Hansen, and Whitney Newey.
\newblock Double/debiased/neyman machine learning of treatment effects.
\newblock {\em American Economic Review}, 107(5):261--265, 2017.

\bibitem{rubin2006matched}
Donald~B Rubin.
\newblock {\em Matched sampling for causal effects}.
\newblock Cambridge University Press, 2006.

\bibitem{robins2000marginal}
James~M Robins, Miguel~Angel Hernan, and Babette Brumback.
\newblock Marginal structural models and causal inference in epidemiology, 2000.

\bibitem{hahn1998role}
Jinyong Hahn.
\newblock On the role of the propensity score in efficient semiparametric estimation of average treatment effects.
\newblock {\em Econometrica}, pages 315--331, 1998.

\bibitem{athey2016recursive}
Susan Athey and Guido Imbens.
\newblock Recursive partitioning for heterogeneous causal effects.
\newblock {\em Proceedings of the National Academy of Sciences}, 113(27):7353--7360, 2016.

\bibitem{wager2018estimation}
Stefan Wager and Susan Athey.
\newblock Estimation and inference of heterogeneous treatment effects using random forests.
\newblock {\em Journal of the American Statistical Association}, 113(523):1228--1242, 2018.

\bibitem{ctx36517209530001081}
Joel~L Horowitz.
\newblock {\em The Bootstrap}, volume~5 of {\em Handbook of Econometrics}.
\newblock North-Holland Pub. Co. ;, Amsterdam ; New York : New York, N.Y. :, 2001.

\bibitem{Romano_2012}
Joseph~P. Romano and Azeem~M. Shaikh.
\newblock On the uniform asymptotic validity of subsampling and the bootstrap.
\newblock {\em The Annals of Statistics}, 40(6), December 2012.

\bibitem{tibshirani1993introduction}
Robert~J Tibshirani and Bradley Efron.
\newblock An introduction to the bootstrap.
\newblock {\em Monographs on statistics and applied probability}, 57(1):1--436, 1993.

\bibitem{word1990state}
Elizabeth Word, John Johnston, Helen~P Bain, B~DeWayne Fulton, Jayne~B Zaharias, Charles~M Achilles, Martha~N Lintz, John Folger, and Carolyn Breda.
\newblock The state of tennessee’s student/teacher achievement ratio (star) project.
\newblock {\em Tennessee Board of Education}, 1990.

\bibitem{kleiber2008applied}
Christian Kleiber and Achim Zeileis.
\newblock {\em Applied Econometrics with R}.
\newblock Springer, 2008.

\bibitem{krueger1999experimental}
Alan~B Krueger.
\newblock Experimental estimates of education production functions.
\newblock {\em The quarterly journal of economics}, 114(2):497--532, 1999.

\bibitem{Cinelli2020}
Carlos Cinelli, Andrew Forney, and Judea Pearl.
\newblock A crash course in good and bad controls.
\newblock {\em SSRN Electronic Journal}, 01 2020.

\end{thebibliography}

\end{document}